\documentclass{elsarticle}

\usepackage[english]{babel} 
\usepackage[utf8]{inputenc}
\usepackage{amsmath}
\usepackage{graphicx}
\usepackage{wrapfig}
\usepackage{hyperref}
\usepackage{sidecap} 
\usepackage[colorinlistoftodos]{todonotes}

\biboptions{sort&compress}

\usepackage{listings}

\definecolor{dkgreen}{rgb}{0,0.6,0}
\definecolor{gray}{rgb}{0.5,0.5,0.5}
\definecolor{mauve}{rgb}{0.58,0,0.82}

\newcommand{\bit}{\begin{itemize}}
\newcommand{\eit}{\end{itemize}}
\newcommand{\be}{\begin{equation}}
\newcommand{\ee}{\end{equation}}

\journal{Computer Physics Communications}

\begin{document}

\begin{frontmatter}

\title{\texttt{GraphState} -- a tool for graph identification and labelling}

\author[ff]{D.~Batkovich}
\ead{batya239@gmail.com}

\author[ff]{Yu.~Kirienko}
\ead{j.kirienko@spbu.ru}

\author[ff]{M.~Kompaniets\corref{cor}}
\ead{mkompan@gmail.com}

\author[ggl]{S.~Novikov}
\ead{dr.snov@gmail.com}

\address[ff]{Saint Petersburg State University, 198504, Russia, Saint Petersburg, Peterhof, Ulianovskaya str., 1}
\address[ggl]{Google, Switzerland}

\cortext[cor]{Corresponding author}

\begin{abstract}
We present python libraries for Feynman graphs manipulation. The key feature of these libraries is usage of generalization of graph representation offered by B. G. Nickel et al. In this approach graph is represented in some unique 'canonical' form that depends only on its combinatorial type. The uniqueness of graph representation gives an efficient way for isomorphism finding, searching for subgraphs and other graph manipulation tasks. Though offered libraries were originally designed for Feynman graphs, they might be useful for more general graph problems.
\end{abstract}

\begin{keyword}
graph isomorphisms\sep Feynman diagrams \sep graph manipulation tools  \sep subgraph search
\end{keyword}
\end{frontmatter}
{\bf PROGRAM SUMMARY}                         
                                                                          
\begin{small}                                                             
\noindent                                                                 
{\em Manuscript Title:}  \texttt{GraphState} -- a tool for graph identification and labelling                                     \\          
{\em Authors:} D.~Batkovich, Yu.~Kirienko, M.~Kompaniets, S.~Novikov                                                 \\          
{\em Program Title:} \texttt{GraphState}, \texttt{Graphine}                                         \\          
{\em Journal Reference:}                                      \\          
{\em Catalogue identifier:}                                   \\          
{\em Licensing provisions:}                                   \\          
{\em Programming language:} \texttt{Python} (2.6, 2.7)                                  \\          
{\em Computer:} Any                                              \\          
{\em Operating system:} Unix-like, Windows                                      \\          
{\em RAM:} 128 Mb                                              \\          
{\em Number of processors used:}                              \\          
{\em Supplementary material:}                                 \\          
{\em Keywords:} graph isomorphisms problem, Feynman diagrams, graphs manipulation tools, subgraphs searching  \\         
{\em Classification:}\\
{\em External routines/libraries:}                            \\
{\em Subprograms used:}                                       \\
{\em Nature of problem:} manipulations with graphs and graphs automorphisms problem
   \\
{\em Solution method:} Extension and implementation of graph labelling introduced by B.G.~Nickel et al.
   \\
{\em Restrictions:} only connected graphs supported
   \\
{\em Unusual features:} graph automorphisms searching and subgraphs searching, 
   \\
{\em Additional comments:}
   \\
{\em Running time:} depends on problem type
   \\

\end{small}

\section{Introduction}
\subsection{Problem Statement}
Dealing with graphs in computer science we face the problem of recognising similar graphs among the set somehow internally represented ones. 
To solve this problem, we need to learn how to solve a number of simpler subproblems: naming and searching of graph isomorphisms. These can be solved by constructing of canonical representation of the graph.

\section{Nickel index}

\subsection{Main principle of constructing indices}

An arbitrary graph can be represented as adjacency matrix (or adjacency list), ordering does not matter there. But in order to construct the canonical representation we need {\em some} ordering. We introduce the following rules for this purpose (the reasons for these rules will become evident later on): 
\bit
\item vertex ordering
\item edges ordering (in sense of "adjacency") 
\item edges "begin-end" ordering (in sense of "where this edge begins from")
\eit


\subsection{Undirected graph labelling and ordering}

The labelling algorithm described in this section was introduced by B.G. Nickel {\em et al.} (based on~\cite{Nagle1966}). It is used to describe connected undirected graphs with “simple“ edges and vertices. Despite of the simplicity, this way of labelling is not widely known, so here we present its main rules. For more details and motivation one can see original paper~\cite{Nickel1977}, where the first $6$-loop results of $\phi^4$-theory were published.

In paper ~\cite{Nickel1977} the notation for undirected graphs  was presented (we call it {\it Nickel notation}). This notation can be defined as follows.
Consider an arbitrary undirected connected graph with $n$ vertices. Nickel notation of this graph can be obtained by labelling the vertices by the integers $0$ through $n-1$  and then constructing the sequence -- a list of vertex lists, according to \cite{Nickel1977}:
\be
\begin{array}{l}
\mbox{vertices connected to vertex 0 $|$ }
\mbox{vertices connected to $1$ excluding $0$ $|$}\\ 
\mbox{vertices connected to $2$ excluding $0, 1$ $|$ \ldots $|$}\\
\mbox{vertices connected to $m$ excluding $0, 1, \ldots, m-1$ $|$ \ldots $|$}\\
\mbox{vertices connected to $n-1$ excluding $0, 1, \ldots, n-2$.}
\end{array}
\label{notation} 
\ee

The description (\ref{notation}) allows relatively compact string representation: one character per graph's edge plus list separators. In our representation we use $0, 1, \dots, 9, A, B, C, \dots$ to label  vertices and "pipe" $|$ as a vertex separator. 

\begin{SCfigure}
	\includegraphics{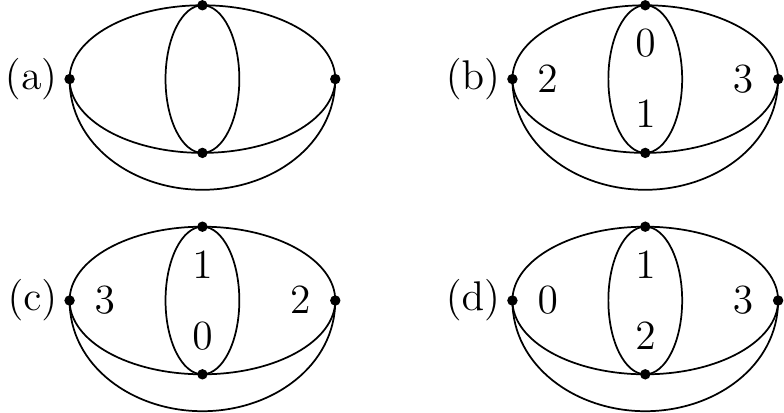}
    \caption{Unenumerated graph (a) can be represented in different forms: there are $4!$ ways of vertex enumeration, and some of them might be isomorphic to each other (b and c) but others are not (c and d)}
    \label{pic_intro}
\end{SCfigure}

Every edge connecting two distinct vertices appears once in (\ref{notation}) and we also adopt the convention that any edge connecting a vertex to itself be listed only once. Nevertheless, if we have more than one line between a pair of vertices, then we should indicate it in the sequence corresponding number of times. (For example, in Fig.~\ref{pic_intro} (upper right) there are two edges between vertex $0$ and vertex $1$, and we need to write $11|$ in the first part.) Using this rule, at the end of enumeration procedure we have every line of graph exactly once in all sequence.
For the subsequent choice of the only «canonical» description it is essential to sort each vertex list in ascending order.

With given description (\ref{notation}), one can reconstruct the graph, but in general there are $n!$ ways of possible labellings, some of them might lead to distinct sequences. For numerated graphs presented in Fig.~\ref{pic_intro}~we got the following Nickel notations: (b) and (c) : $1123|23|3||$, (d) : $123|223|3||$. 

To describe unenumerated graph we need to construct some description which depends only on combinatorial type of graph and does not depend on vertex numeration, we call this description as {\it Nickel index}, it can be constructed 
by finding "minimum" among the Nickel notations using following procedure:
\begin{enumerate}
\item Get the set of all possible labellings of graph's vertices.
\item For each labelling obtain description according to~\eqref{notation}.
\item Order the set of descriptions lexicographically, i.e. according to the following rule of comparison: compare pairs of corresponding vertex lists from both descriptions one-by-one from vertex $0$ to $n-1$. If we found a pair in which one list precedes another in lexical order then the corresponding description is considered to be “less” than another one.
\item Choose the first description (i.e. {\em minimal} in this sense) as {\em Nickel index}.
\end{enumerate}
\begin{figure}[h!]
\centering
\includegraphics{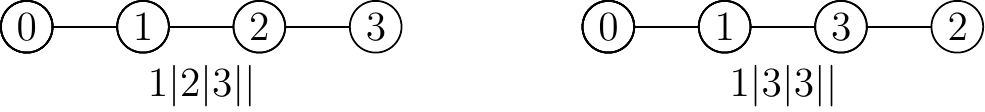}
\caption{\label{fig:path_graph} Different enumerations produce different descriptions}
\end{figure}
In other words {\em Nickel index} is a unique graph representation with the use of {\em Nickel notation}.

For example (see Fig.~\ref{fig:path_graph}) description $1|2|3||$ is less than $1|3|3||$ as 2 precedes 3 in lexical order.

The algorithm internally uses bread first search with pruning of non-minimal partial Nickel notation. This way it reduces $O(n!)$ complexity of original problem to $O(n)$ for most cases. The worst case is fully connected graph for which the algorithm degenerates back to $O(n!)$.

\subsection{External legs}

Wide class of problems in theoretical physics~\cite{Vasilev2004, Zinn2007} requires the introduction of special kind of mathematical objects -- the so-called {\em amputated Green function}, that can be represented via {\em Feynman diagrams}, which are in fact graphs with external legs. 

External edges are regarded as terminating on a vertex labelled $e$. In Nickel index we always can put external edge first in the list of edges. Every "internal" vertex supposed to be lexicographically "bigger" than $e$.

\begin{figure}[h!]
\centering
\includegraphics{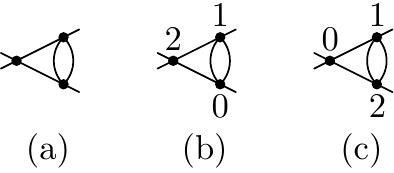}
\caption{\label{fig:eye} For different vertex enumerations one can obtain different representations: $e112|e2|ee|$ (centre) and $ee12|e22|e|$ (right)}
\end{figure}

	For unenumerated graph on Fig.~\ref{fig:eye}(a) "canonical" vertex ordering is~\ref{fig:eye}(c), that corresponds to Nickel index $ee12|e22|e|$. 

\subsection{Generalized case}

Things above was used by Nickel for calculation of $\phi^4$ in spaces of fixed dimension. In more complicated theories~\cite{Bednyakov2014} Feynman diagrams can not be presented only by undirected graphs. We should generalise out the description. We might need vertices and edges of different type or directed edges. 
In this subsection we describe generalisaton of Nickel indices for such a case.

Nickel index fixes ordering: firstly, it is an ordering of vertices, secondly, for ordered vertices it canonically orders sequence of edges, and finally, it fixes the directions of all the edges (by agreement, all edges directed from vertices with less number to the vertices with bigger number). This fact allows us to extend Nickel notation to the wider class of theories by noting properties of vertices and edges according to the ordering mentioned.

If graph possesses symmetries, then we have several ways of vertex ordering that produce the same Nickel index. Then canonical ordering is that minimizes extended Nickel index according to additional properties. In this subsection we will give examples of extended Nickel notation.

\subsubsection{Coloured vertices [with markers]}

\begin{wrapfigure}{r}{0.25\textwidth}
\includegraphics{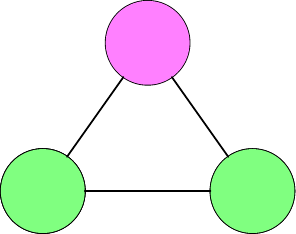}
\caption{Coloured vertices}
\label{fig:triangle}
\end{wrapfigure}

Some models demand existence of different properties of vertices with the same vertex degree. We can associate different markers with these vertices, and refer such a vertices as "coloured". 

As soon as Nickel index fixes vertex ordering, we can count the colours of vertices in that order.
It can be done by adding additional section separated from the main one by ":", in this section vertex properties are listed in the order defined by first section and also separated from each other by "$|$".

Example: Triangle graph with two vertices of type $a$ (green) and one veretex of type $b$ (pink) (Fig.~\ref{fig:triangle}).

For uncoloured case we have 3 equivalent ways of ordering which lead to minimal Nickel notation. For each ordering we add to Nickel notation additional section with vertex properties. This results in the following three notations: $12|2||:a|a|b|$, $12|2||:a|b|a|$ and $12|2||:b|a|a|$. Now to construct Nickel index we need find minimal notation, this assumes that there is some {\em ordering relation on vertex properties(colours)}, e.g. $a<b<c<\dots$. This relation allows one to choose one of this notations as Nickel index, in our case minimal one is $12|2||:a|a|b|$

\subsubsection{Edges of different type}
\begin{wrapfigure}{r}{0.25\textwidth}
\includegraphics{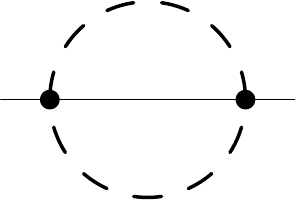}
\caption{Coloured lines}
\label{fig:arbuz}
\end{wrapfigure}

Exactly the same way as in the previous section, we can define coloured edges: as soon as Nickel index fixes edges ordering, we can count the colours of edges in that order. Therefore we can equip Nickel notation with additional part, that describe properties of edges.

Particular order of different extensions of notation is the question of agreement, depending on underlying problem to solve. For example, we can say that coloured lines precede coloured vertex in the notation, if for some reason we assume coloured properties of edges as more "valuable" for our purposes.

Example: Consider sunset diagram (Fig.~\ref{fig:arbuz}) with two types of lines: non-dashed 0 and dashed 1. Then possible Nickel notations of diagram can be $e111|e|:0\_0\_0\_1|0|$, $e111|e|:0\_1\_0\_0|0|$ or $e111|e|:0\_0\_1\_0|0|$. First notation is minimal, then this is Nickel index of given diagram.

\subsubsection{Directed graphs}
\begin{wrapfigure}{r}{0.25\textwidth}
\includegraphics{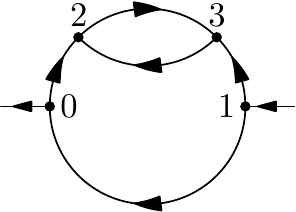}
\caption{Directed lines}
\label{fig:directed}
\end{wrapfigure}
Since Nickel index fixes direction of every edge (from less vertex number to bigger), we can extend the notation with a part, that describes whether every edge has that "canonical" direction or it is directed oppositely.

Simple directed graphs can be described using following convention: let us denote by "$>$" the edge directed in the same way as "canonical" direction and by "$<$" if direction is opposite. And to introduce some ordering on this property we assume that "$<$" is less than "$>$"

Example. Following rules above graph on Fig.~\ref{fig:directed} has the following Nickel index
$e12|e3|33||:{<}\,\_{<}\,\_\,{>}|{>}\_\;{>}|{<}\,\_\,{>}||$

\subsection{Put everything together}
We can think about any arbitrary complex graph in a way of layered structure: at the bottom level there are "bones", undirected graph with no special properties. Then we can lay the next "layer" on top: edges direction. Providing this procedure step by step, increasing the complexity of naming with every step, we can fully depict our graph. 

The order of definition of the graph properties matters, because it leads to different Nickel indices (two isomorphic graphs with different order of property enumeration have different Nickel indices).
In general, one should place more important properties earlier.

Suppose we have a set of diagrams with both coloured vertices and coloured edges. And in some certain case we suppose the edges' properties as more important. Then all diagrams, that possess the same first section, have the same topology; all diagrams with the same first and second sections, share similar topology and edges properties.

Also this rule of enumeration allows to group diagrams in the most convenient way. I.e. if one decide to put the properties of vertices before the properties of edges, in such a case diagrams become sorted according to topology and vertex properties first.

\section{GraphState}

In the following section we provide multiple examples of code that illustrate {\tt GraphState} capabilities and features.

\subsection{Configure}
In previous section we presented the way of describing different types of graphs. 

If we want to set any special property on a graph, we should define what kind of properties and in which order we are going to use.
This can be done as follow:
\begin{lstlisting}[language=Python]
import graph_state

config = graph_state.PropertiesConfig.create(*property_keys)
\end{lstlisting}
where {\tt *property\_keys} defines edges' and vertices' properties, its behaviour and serialization/deserialization rules (see~\ref{graphState_properties}).
It might be useful to store the code that creates property configuration in separate file in order to reuse it in different programs.

\subsection{Graphs with simple edges}

To describe the simple graph without properties, one should not pass any arguments to {\tt PropertiesConfig.create()}.
\begin{lstlisting}[language=Python]
>>> import graph_state
>>> simple_config = graph_state.PropertiesConfig.create()
\end{lstlisting}

\begin{figure}
\centering
\includegraphics{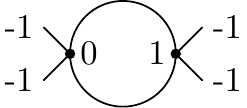}
\caption{Diagram with Nickel index $ee11|ee|$. Vertices are labelled as $0$ and $1$; each 'external' edges begin at vertex, conventionally labelled as $-1$.}
\label{fig:bubble}
\end{figure}

For example, we can create a simple diagram (see Fig~\ref{fig:bubble}) from the list of edges (-1 means that the edge is external):

\begin{lstlisting}[language=Python]
>>> edges = [(1, -1), (1, -1), (-1, 0), (-1, 0), (0, 1), (0, 1)]
>>> graph_state_edges = map(simple_config.new_edge, edges)
>>> bubble = simple_config.new_graph_state(graph_state_edges)
>>> bubble
ee11|ee|
\end{lstlisting}

The same result can be achieved by getting the state of the graph (via Nickel notation) from string:
\begin{lstlisting}[language=Python]
>>> another_bubble = simple_config.graph_state_from_str('ee11|ee|')
ee11|ee|
\end{lstlisting}

If non minimal Nickel notation is passed as argument of method  \texttt{graph\_state\_from\_str()}, then minimal Nickel notation (i.e. Nickel index) will be constructed.

\texttt{GraphState} object has following property to access nodes of graph: 

\begin{lstlisting}
>>> bubble.nodes
[-1, 0, 1]
>>> bubble.nodes[0] == -1
True
>>> bubble.nodes[0] is -1
False
>>> type(bubble.nodes[0])
<class 'graph_state.graph_state_property.Node'>
\end{lstlisting}

Information about edges of graph can be accessed by following ways:

\begin{lstlisting}
>>> bubble.edges
(((0,), ), ((0,), ), ((0, 1), ), ((0, 1), ), ((1,), ), ((1,), ))
>>> [e for e in bubble] # iter through edges
[((0,), ), ((0,), ), ((0, 1), ), ((0, 1), ), ((1,), ), ((1,), )]
>>> len(bubble) # count of edges
6
\end{lstlisting}

Every edge is represented as \texttt{graph\_state.Edge} object, and you can call its methods and properties to access nodes from single edge:

\begin{lstlisting}
>>> isintance(another_bubble.edges[0], graph_state.Edge)
True
>>> another_bubble.edges[0].nodes
(-1, 0)
>>> another_bubble.edges[0].internal_node
0
>>> another_bubble.edges[0].internal_nodes
(0,)
>>> another_bubble.edges[0].external_node
-1
>>> another_bubble.edges[0].co_node(-1)
0
>>> another_bubble.edges[0].co_node(0)
-1
>>> another_bubble.edges[0].is_external
True
\end{lstlisting}

Here \texttt{(0, )} means that external edge starts from vertex 0.

\subsection{\label{graphState_properties} Graphs with properties}

To create edges or vertices with some properties, we need to define non-trivial PropertiesConfig. Property can be any object with defined \texttt{\_\_cmp\_\_} method.

\begin{lstlisting}
>>> graph_state.PropertiesConfig.create(*property_keys)
\end{lstlisting}

Snippet above shows how \texttt{property\_keys} define edge or vertex properties. The order of keys will define the order of properties' "layers".

\begin{lstlisting}
graph_state_property.PropertyKey(name, is_directed=False, is_edge_property=True, externalizer=graph_state.FakePropertyExternalizer())
\end{lstlisting}

In this snippet externalizer object serves as rule to define serialization to string and deserialization from string of given property object. Any externalizer must inherits \texttt{PropertyExternalizer} class and overrides two methods described below. 

\begin{lstlisting}
class CustomPropertyExternalizer(PropertyExternalizer):
    def deserialize(self, string):
        # object building from given string
        return built_object

    def serialize(self, obj):
        # string building from object obj
        return built_str
\end{lstlisting}

By default \texttt{FakePropertyExternalizer} is used. This externalizer just raises \texttt{NotImplementedExceptions} from both methods. Additionally out of the box we have \texttt{graph\_state.StringPropertyExternalizer} externalizer which deserializes string by evaluating it with the function \texttt{eval} and serializes using \texttt{str} function.

Property will be injected to edge or vertex as python property with name equals to \texttt{name} parameter of \texttt{PropertyKey}, for example to access edge property with name "weight" you need just call \texttt{edge.weight} 

\subsubsection{Coloured vertices}

Consider graph with two types of vertices -- "a" and "b" (any string value can also be approached). Correspondent config is given by following code:

\begin{lstlisting}
>>> vertex_property_key = graph_state_property.PropertyKey(name="color", is_edge_property=False, externalizer=graph_state.StringPropertyExternalizer())
>>> coloured_vertices_config = graph_state.PropertiesConfig.create(vertex_property_config)
\end{lstlisting}

Then triangle graph can be created:

\begin{lstlisting}
>>> coloured_triangle = coloured_vertices_config.graph_state_from_str("12|2||:a|a|b|")
>>> coloured_triangle.vertices[0].color == 'a'
True
\end{lstlisting}

\subsubsection{Coloured lines}

Consider graph with two types of edges - 0 and 1. Correspondent config is given by following code:

\begin{lstlisting}
>>> edge_property_key = graph_state_property.PropertyKey(name="color", is_edge_property=True, is_directed=False, externalizer=graph_state.PropertyExternalizer())
>>> coloured_edges_config = graph_state.PropertiesConfig.create(vertex_property_config)
\end{lstlisting}

Then sunset diagram can be created:

\begin{lstlisting}
>>> coloured_sunset = coloured_edges_config.graph_state_from_str("e111|e|:0_0_0_1|0|")
>>> coloured_sunset.edges[0].color is 0
True
\end{lstlisting}

\subsubsection{Directed graphs}

In last example with custom properties we demonstrate all power of externalizers. Assume that we have some class with name \texttt{Arrow}. It has factory static method \texttt{Arrow.from\_str(some\_string)}, where argument can takes one of three values -- \texttt{>}, \texttt{<} or \texttt{0}. And exception will be thrown in any other case. Additionally assume that method \texttt{Arrow.\_\_str\_\_} returns strictly string that was a parameter of factory method. And finally methods \texttt{Arrow.is\_null}, \texttt{Arrow.is\_left}, \texttt{Arrow.is\_right} defined naturally. Now we define externalizer for this object.

\begin{lstlisting}
class ArrowExternalizer(graph_state_property.PropertyExternalizer):
    def deserialize(self, string):
        return Arrow.create(string)
    
    def serialize(self, obj):
        return str(obj)
\end{lstlisting}

Then create bubble with one arrow

\begin{lstlisting}
>>> arrow_key = graph_state_property.PropertyKey(name="arrow", is_edge_property=True, is_directed=True, externalizer=ArrowExternalizer())
>>> arrowed_edges_config = graph_state.PropertiesConfig.create(arrow_key)
\end{lstlisting}

\begin{lstlisting}
>>> bubble_with_arrow = arrowed_edges_config.graph_state_from_str("e11|e|:0_>_<|0|")
>>> bubble_with_arrow.edges[0].arrow.is_null()
True
>>> bubble_with_arrow.edges[1].arrow.is_right()
True
>>> bubble_with_arrow.edges[2].arrow.is_left()
True
>>> str(bubble_with_arrow)
e11|e|:0_>_<|0|
\end{lstlisting}

\subsection{Auxiliary possibilities}

Auxiliary we provide basic graph-theoretical operations for \texttt{GraphState} objects or list of edges. 

\begin{lstlisting}
>>> from graph_state import operations_lib
>>> help(operations_lib)
\end{lstlisting}

Library provides getting connected components, checking that graph is connected, 1-irreducible, vertex irreducible and some other useful util functions. All list of features you can see at \texttt{GraphState/graph\_state/operations\_lib.py}. Note that for functions which is marked with \texttt{@graph\_state\_to\_edges\_implicit\_conversion} first parameter can both list of edges and \texttt{GraphState} types. 

\section{\texttt{Graphine}}

\texttt{Graphine} provides an extended representation of the graph which allows easy access to graph structure, modifying graphs and searching for subgraphs defined by some condition. Additionally \texttt{Graphine} represents graph as immutable object, and results of all methods cached in internal graph cache. Class \texttt{graphine.Graph} is used as the graph representation. \texttt{graphine.Graph} is short life object because it contains strong reference cache in backend, hence memory leaks are possible.

There are two ways to create \texttt{Graph} object from string or from already existing \texttt{GraphState} object:

\begin{lstlisting}
>>> from graphine import Graph
>>> Graph.from_str("e11|e|", your_graph_state_config)
>>> Graph(your_graph_state_object)
\end{lstlisting}
To access graph structure
\begin{lstlisting}
graph.vertices
graph.external_edges
graph.internal_edges
graph.loops_count
\end{lstlisting}
can be used or:
\begin{lstlisting}
>>> for e in graph:
      do_some(e)
>>> for e in graph.edges(v1):
      # edges which have v1 as vertex
      do_some(e)
>>> for e in graph.edges(v1, v2):
      # edges which connect directly v1 and v2 vertices
      do_some(e)
\end{lstlisting}
To immutably modify graph following methods can be used:
\begin{lstlisting}
graph_with_shrunk_subgraph = graph.shrink_to_point(sub_graph)
graph_with_shrunk_subgraphs = graph.batch_shrink_to_point(sub_graphs)
modified_graph = graph.change(edges_to_delete, edges_to_add)
\end{lstlisting}

Last method \texttt{change} deletes \texttt{edges\_to\_delete} and adds \texttt{edges\_to\_add} from initial graph edges and then creates new graph.

\subsection{Searching for subgraphs defined by some condition}
Sometimes it is useful to find subgraphs of initial graph which satisfy the given conditions. To determine, whether the subgraph fits given conditions, we use {\em graph filters}. Filter is function of 2 arguments -- \texttt{sub\_graph\_edges} and \texttt{super\_graph} with added decorator \texttt{graphine.filters.graph\_filter}.

Some of useful filters are predefined:
\begin{lstlisting}
graphine.filters.one_irreducible  # one particle irreducible
graphine.filters.connected
graphine.filters.no_tadpoles  # no tadpoles in co_subgraph
graphine.filters.vertex_irreducible
\end{lstlisting}
To create composite filter just use \texttt{+}:
\begin{lstlisting}
graph.x_relevant_sub_graphs(filters=connected + one_irreducible)
\end{lstlisting}
The result will be an iterator of subgraphs which satisfy given filter.

In example below one can find the filter for searching for $UV$-divergent subgraphs in $\phi^{4}$ model. It is implemented by simple $UV$ power counting and must be combined with at least \texttt{one\_irreducible} filter.

\begin{lstlisting}
@graphine.filters.graph_filter
def phi4_uv_subgraph_filter(self, edges_list, super_graph):
    sub_graph = graphine.Representator.asGraph(edges_list)
    n_edges = len(sub_graph.internal_edges)
    n_loop = sub_graph.loops_count
    sub_graph_uv_index = -2 * n_edges  + 4 * n_loop
    return sub_graph_uv_index >= 0
\end{lstlisting}

\section{Installation}

To install \texttt{Graphine} and \texttt{GraphState} libraries you need to download tarballs from \href{https://code.google.com/p/rg-graph/}{https://code.google.com/p/rg-graph/}. After it has been downloaded go to the download folder and execute following commands from CLI to install \texttt{GraphState}:

\begin{lstlisting}
$ tar xvfz GraphState.tar.gz
$ cd GraphState
$ python setup.py install --user
\end{lstlisting}

and after these steps do similar commands for \texttt{Graphine} library.

\section*{Acknowledgements}
The authors acknowledge Saint-Petersburg State University for a research grant 11.38.185.2014. Authors would like to
thank {\em L.Ts.~Adzhemyan} for fruitful discussions during development of the library and {\it A.V.~Bednyakov} for valuable feedback after using \texttt{GraphState} library.

\bibliographystyle{plain}
\bibliography{main}
\end{document}